\documentclass[a4paper,12pt]{article}
\usepackage{amsfonts}
\usepackage{latexsym}
\usepackage{amssymb}
\date{}

\begin{document}

\title{Entropy Production and Equilibrium Conditions of\\
General-Covariant Spin Systems} 
\author{W. Muschik\footnote{Corresponding author:
muschik@physik.tu-berlin.de}\quad and\quad 
H.-H. von Borzeszkowski\footnote{ 
borzeszk@mailbox.tu-berlin.de}
\\
Institut f\"ur Theoretische Physik\\
Technische Universit\"at Berlin\\
Hardenbergstr. 36\\D-10623 BERLIN,  Germany}
\maketitle

            \newcommand{\be}{\begin{equation}}
            \newcommand{\beg}[1]{\begin{equation}\label{#1}}
            \newcommand{\ee}{\end{equation}\normalsize}
            \newcommand{\bee}[1]{\begin{equation}\label{#1}}
            \newcommand{\bey}{\begin{eqnarray}}
            \newcommand{\byy}[1]{\begin{eqnarray}\label{#1}}
            \newcommand{\eey}{\end{eqnarray}\normalsize}
            \newcommand{\beo}{\begin{eqnarray}\normalsize}
            \newcommand{\R}[1]{(\ref{#1})}
            \newcommand{\C}[1]{\cite{#1}}

            \newcommand{\mvec}[1]{\mbox{\boldmath{$#1$}}}
            \newcommand{\x}{(\!\mvec{x}, t)}
            \newcommand{\m}{\mvec{m}}
            \newcommand{\F}{{\cal F}}
            \newcommand{\n}{\mvec{n}}
            \newcommand{\argm}{(\m ,\mvec{x}, t)}
            \newcommand{\argn}{(\n ,\mvec{x}, t)}
            \newcommand{\T}[1]{\widetilde{#1}}
            \newcommand{\U}[1]{\underline{#1}}
            \newcommand{\X}{\!\mvec{X} (\cdot)}
            \newcommand{\cd}{(\cdot)}
            \newcommand{\Q}{\mbox{\bf Q}}
            \newcommand{\p}{\partial_t}
            \newcommand{\z}{\!\mvec{z}}
            \newcommand{\bu}{\!\mvec{u}}
            \newcommand{\rr}{\!\mvec{r}}
            \newcommand{\w}{\!\mvec{w}}
            \newcommand{\g}{\!\mvec{g}}
            \newcommand{\D}{I\!\!D}
            \newcommand{\se}[1]{_{\mvec{;}#1}}
            \newcommand{\sek}[1]{_{\mvec{;}#1]}}            
            \newcommand{\seb}[1]{_{\mvec{;}#1)}}            
            \newcommand{\ko}[1]{_{\mvec{,}#1}}
            \newcommand{\ab}[1]{_{\mvec{|}#1}}
            \newcommand{\abb}[1]{_{\mvec{||}#1}}
            \newcommand{\td}{{^{\bullet}}}
            \newcommand{\eq}{{_{eq}}}
            \newcommand{\eqo}{{^{eq}}}
            \newcommand{\f}{\varphi}
            \newcommand{\dm}{\diamond\!}
            \newcommand{\seq}{\stackrel{_\bullet}{=}}
            \newcommand{\st}[2]{\stackrel{_#1}{#2}}
            \newcommand{\om}{\Omega}
            \newcommand{\emp}{\emptyset}
            \newcommand{\bt}{\bowtie}
            \newcommand{\btu}{\boxdot}
\newcommand{\Section}[1]{\section{\mbox{}\hspace{-.6cm}.\hspace{.4cm}#1}}
\newcommand{\Subsection}[1]{\subsection{\mbox{}\hspace{-.6cm}.\hspace{.4cm}
\em #1}}

\newcommand{\const}{\textit{const.}}
\newcommand{\vect}[1]{\underline{\ensuremath{#1}}}
\newcommand{\abl}[2]{\ensuremath{\frac{\partial #1}{\partial #2}}}

\abstract\noindent
In generalizing the special-relativistic one-component version of Eckart's
con\-ti\-nuum thermodynamics to general-relativistic space-times with
Riemannian or post-Rie\-mannian geometry\footnote{For Riemannian and post-Riemannian geometries, see \C{SCH}. For a survey on  gravitational
theories based on space-times with post-Riemannian geometry, see \C{BH}}, 
we consider the entropy production
and other themodynamical quantities such as the entropy flux and the
Gibbs fundamental equation. We discuss equilibrium conditions in gravitational
theories which are based on such geo\-metries. In particular,
thermodynamic implications of the non-symmetry of the energy-momentum
tensor and the related spin balance equations are investigated, also for
the special case of General Relativity.

\section{Introduction}

The special-relativistic version of Continuum Thermodynamics (CT) was
founded by Eckart \C{EC} in form of the special-relativistic theory of
irreversible processes. CT is based (i) on the conservation law of the
particle number and on the balance equation of the energy-momentum
tensor and (ii) on the dissipation inequality and the Gibbs fundamental
equation. In order to incorporate CT into
General Relativity (GR) and other gravitational theories all based on
curved space-times, as a first step, one has to go over to the
general-covariant formulation of CT which is performed here.
\vspace{.3cm}\newline
The paper is devoted to derive the entropy production and equilibrium conditions
in General-Covariant Continuum Thermodynamics (GCCT).
Starting out with an entropy identity \C{B} --a tool to
construct entropy flux and production, as well as gr-Gibbs and gr-Gibbs-Duhem
equations more safely-- different forms of the entropy production
are considered for discussing
non-dissipative materials and equilibria which both are characterized by vanishing
entropy production. For defining equilibrium beyond the vanishing entropy
production, additionally ``supplementary equilibrium conditions'' are required \C{B}.
The material-independent equilibrium condition --that the 4-temperature
vector is a Killing field-- is rediscovered also for gravi\-tational theories beyond GR
including a spin part in the state space. 
\vspace{.3cm}\newline
The paper is organized as follows: first the general-covariant shape of the
energy-momentum and spin CT-balances are written down and entropy
flux and production, gr-Gibbs and gr-Gibbs-Duhem equations are derived.
Non-dissipative materials and equilibria of spin materials are investigated
with regard to the resulting constitutive constraints. Finally, equilibrium
conditions for Frenkel materials are derived.

\section{General-Covariant Continuum Physics}

\subsection{The balance equations}

The balance equations of energy-momentum and spin of phenomenological
GCCT in a curved
space-time\footnote{The comma denotes partial and the semicolon covariant
derivatives, round brackets the symmetric part of a tensor, square brackets
its asymmetric part.} are \C{B}
\byy{b5a}
T^{bc}_{\ \ ;b}\ =\ G^c + k^c,\quad T^{bc}\ \neq\ T^{cb}, 
&\quad& S^{cba}{_{;c}}\ =\ H^{ba} +m^{ba},
\\ \label{b5a1}
\mbox{with}\quad S^{cba}\ =\ -S^{cab},\ m^{ba}\ =\ -m^{ab}
&\mbox{and}& H^{ba}\ =\ -H^{ab}.
\eey
Here, $T^{ab}$ is the in general non-symmetric energy-momentum tensor
of CT and
$S^{cba}$ the current of spin density\footnote{often shortly denoted as
spin tensor}. The $G^c$ and $H^{bc}$ are internal source terms
--the {\em Geo-SMEC-terms}\footnote{\U{Geo}metry-\U{S}pin-\U{M}omentum-\U{E}nergy-\U{C}oupling} \C{B}-- which are caused by the choice of a special space time
and by a possible coupling between energy-momentum, spin and geometry.
\vspace{.3cm}\newline
For non-isolated systems, $k^c\neq 0$ denotes an external force
density, and\newline $m^{ab}\neq 0$ is an external momentum density.
As in the continuum theory of irreversible processes \C{I} \C{N},
the balance equations \R{b5a} must be supplemented by
those of particle number and entropy density
\bee{P3}
N^k{_{;k}}\ =\ 0, \qquad S^k{_{;k}}\ =\ \sigma + \varphi
\ee
($N^k$ particle flux density, $S^k$ entropy 4-vector, $\sigma$
entropy production, $\varphi$ entropy supply). The Second Law of
Thermodynamics is taken
into account by the demand that the entropy production has to be non-negative
at each event and for arbitrary materials\footnote{after having inserted the constitutive
equations into the expression of the entropy production}
\bee{P4}
\sigma\ \geq\ 0.
\vspace{.3cm}\ee
The (3+1)-splits of the tensors in \R{b5a} and \R{P3} are
\byy{P8}
N^k &=& \frac{1}{c^2}nu^k,\quad n\ :=\ N^ku_k,\quad (nu^k)_{;k}\ =\ 0,
\\ \label{P2c}
T^{kl} &=& \frac{1}{c^4}eu^ku^l + \frac{1}{c^2}u^kp^l +
\frac{1}{c^2}q^ku^l + t^{kl},
\\ \label{aP2c}
&&p^lu_l\ =\ 0,\quad q^ku_k\ =\ 0,\quad t^{kl}u_k\ =\ 0,\quad
t^{kl}u_l\ =\ 0,
\\ \label{L15b}
S^{kab} &=&
\Big(\frac{1}{c^2}s^{ab} + \frac{2}{c^4}u^{[a}\Xi^{b]}\Big)u^k
+s^{kab} + \frac{2}{c^2}u^{[a}\Xi^{kb]}\ =:\ 
u^k\Phi^{ab} + \Psi^{kab},
\\ \label{L15c}
&&\Xi^bu_b\ =\ 0,\quad\Xi^{kb}u_k\ =\ \Xi^{kb}u_b\ =\ 0,
\quad s^{ab}u_a\ =\ s^{ab}u_b\ =\ 0,
\\ \label{P6}
S^k &=& \frac{1}{c^2}su^k + s^k,\quad s\ :=\ S^ku_ k,\quad s^k\ :=\ h^k_l S^l.
\eey
Here, the divergence-free particle number flux density $N^k$ is chosen
according to Eckart \C{EC}, and the projector perpendicular to the
4-velocity $u^k$, respective $u_i$, is introduced
\bee{P7}
h^i_k\ =\ \delta^i_k - \frac{1}{c^2}u^iu_k, \quad h^i_ku_i\ =\ 0,
\quad h^i_ku^k\ =\ 0,
\ee
and \R{L15c} results in
\bee{P7a}
\Xi^bh_b^j\ =\ \Xi^j,\quad\Xi^{kb}h_k^j\ =\ \Xi^{jb},\quad\Xi^{kb}h_b^j\ =\ \Xi^{kj}.
\vspace{.3cm}\ee
By splitting the stress tensor into its diagonal
and its traceless parts
\bee{L0a}
t^{kl}\ =\ -p h^{kl} + \pi^{kl},\qquad \pi^{kl}h_{kl}\ =\ 0,\quad
t^{kl}h_{kl}\ =\ t^k_k\ =:\ - 3p, 
\ee
we introduce the pressure $p$ and the friction tensor $\pi^{kl}$. 
\vspace{.3cm}\newline
According to \R{P2c} and \R{aP2c}, we obtain
\bee{E1}
u_lT^{kl}\ =\ q^k + \frac{1}{c^2}eu^k.
\ee
Starting out with \R{L15b}, it holds
\bee{E3}
\Xi^c\ =\ S^{abc}u_au_b,\qquad\Xi^{mc}\ =\ S^{abc}h^m_au_b, 
\ee
resulting in
\bee{E4}
\Xi^{mc} + \frac{1}{c^2}\Xi^cu^m\ =\ S^{mbc}u_b.
\ee
Taking \R{P7a} into account, we obtain
\bee{E4a}
S^{mbc}u_bh_c^j\ =\ S^{mbj}u_b.
\ee

\subsection{The entropy identity}

For establishing the entropy balance equation, we use a special procedure
starting out with an identity \C{B}, the so-called {\em entropy identity}.
This results by
multiplying \R{P8}$_1$, \R{E1} and \R{E4} with for the present arbitrary
quantities $\kappa$,
$\lambda$ and $\Lambda_c$ --which are suitably chosen below-- \R{P6}$_1$
becomes
\bey\nonumber
S^k &\equiv& \frac{1}{c^2}su^k + s^k + \kappa\Big[N^k-\frac{1}{c^2}nu^k\Big]+
\\ \label{aE5}
&&+\ \lambda\Big[u_lT^{kl}-q^k -\frac{1}{c^2}eu^k\Big]\ +\ 
\Lambda_c\Big[S^{kbc}u_b - \Xi^{kc} - \frac{1}{c^2}\Xi^cu^k\Big]\ =\
\\ \nonumber
&=& \frac{1}{c^2}su^k -\kappa\frac{1}{c^2}nu^k-\lambda\frac{1}{c^2}eu^k
 - \frac{1}{c^2}\Lambda_c \Xi^cu^k +
\\ \label{E5}
&&+\ \kappa N^k + \lambda u_lT^{kl}
+ \Lambda_c S^{kbc}u_b +
\Big(s^k - \lambda q^k - \Lambda_c \Xi^{kc}\Big).
\vspace{.3cm}\eey
This entropy identity does not take the entire energy-momentum and spin tensor
into account, but only their contractions with $u_l$ according to \R{aE5}.
This means, that the contractions with $h_{kl}$ are not included in the entropy
identity resulting in the consequence that more than one entropy
identity can be established, if other secondary conditions are taken into
consideration. Consequently, also the results received by exploiting the entropy
identity are not unique, but depend on the chosen entropy identity. Here, we
start out with \R{aE5}.
\vspace{.3cm}\newline
Because the part of $\Lambda_c$ which is parallel to $u^c$ does not
contribute to the last term of \R{aE5} --and consequently not to the
entropy identity-- we can demand 
\bee{bE5}
\Lambda_cu^c\ \doteq\ 0
\ee
without restricting the generality.
The identity \R{E5} becomes another one by differentiation
\bey\nonumber
S^k{_{;k}} &\equiv&
\Big[\frac{1}{c^2}\Big(s-\kappa n-\lambda e-
\Lambda_c \Xi^c \Big)u^k\Big]_{;k}+
\\ \nonumber
&&+\ \Big(\kappa N^k\Big)_{;k} +
\Big(\lambda u_mT^{km}\Big)_{;k}+\Big(\Lambda_c S^{kbc}u_b\Big)_{;k}+
\\ \label{E6}
&&+\ \Big(s^k - \lambda q^k - \Lambda_c \Xi^{kc}\Big)_{;k}.
\eey
This identity changes into the entropy production, if according to \R{P3}$_2$
and \R{P6}$_1$, $s,\ s^k$ and $\varphi$ are specified. For achieving that, we now transform
the five terms of \R{E6}.
\vspace{.3cm}\newline
Introducing the time derivative\footnote{$\ \st{\td}{\boxplus}:=\boxplus_ku^k$ is the
relativistic analogue to the
non-relativistic material time derivative $d\boxplus/dt$ which describes the time
rates of a rest-observer. Therefore, $\st{\td}{\boxplus}$ is observer-independent
and zero in equilibrium \C{D} \C{E}.} and implying the balance equations 
\R{P3}$_1$ and \R{b5a}$_{1,3}$, the entropy identity \R{E6} becomes
\bey\nonumber
S^k{_{;k}} &\equiv&
\frac{1}{c^2}\Big(\st{\td}{s}-\st{\td}{\kappa}n-\kappa\st{\td}{n}-
\st{\td}{\lambda}e-\lambda\st{\td}{e}-\st{\td}{\Lambda}_c\Xi^c -
\Lambda_c\st{\td}{\Xi}{^c}\Big) +
\\ \nonumber
&&+\ \frac{1}{c^2}\Big(s-\kappa n-\lambda e-
\Lambda_c \Xi^c \Big)u^k{_{;k}} +\kappa_{;k}\frac{1}{c^2}nu^k + 
\\ \nonumber
&&+\ \Big(\lambda u_m\Big)_{;k}T^{km} +
\lambda u_m(G^m + k^m) + 
\\ \nonumber
&&+\ \Big(\Lambda_cu_b\Big)_{;k} S^{kbc}+\Lambda_cu_b(H^{bc}+m^{bc})+
\\ \label{E7}
&&+\  \Big(s^k - \lambda q^k - \Lambda_c \Xi^{kc}\Big)_{;k}.
\vspace{.3cm}\eey
Taking \R{E1}, \R{P2c} and \R{L0a} into account, the fourth term
of \R{E7} becomes
\bey\nonumber
\Big(\lambda u_m\Big)_{;k}T^{km}&=&
\lambda_{,k}\Big(q^k +\frac{1}{c^2}eu^k \Big) + 
\lambda u_{m;k}\Big(\frac{1}{c^2}u^kp^m -ph^{km} + \pi^{km}\Big)\ =
\\ \label{P13}
&=&\lambda_{,k}q^k + \frac{1}{c^2}\st{\td}{\lambda}e +
\frac{1}{c^2}\lambda\st{\td}{u}_mp^m - \lambda pu^k{_{;k}}+
\lambda u_{m;k}\pi^{km}.
\vspace{.3cm}\eey
We now transform the sixth term of the entropy identity \R{E7} by taking
\R{L15b} and \R{bE5} into account
\bey\nonumber
\Big(\Lambda_cu_b\Big)_{;k}S^{kbc} &=&
\Big(\Lambda_cu_b\Big){^\td}\Phi^{bc}+
\Big(\Lambda_cu_b\Big)_{;k}\Psi^{kbc}\ =
\\ \nonumber
&=& \Big(\st{\td}{\Lambda}_c u_b+\Lambda_c\st{\td}{u}_b\Big)
\Big(\frac{1}{c^2}s^{bc}+\frac{1}{c^4}(u^b\Xi^c-u^c\Xi^b)\Big) +
\\ \nonumber
&&+\ \Big(\Lambda_{c;k}u_b + \Lambda_cu_{b;k}\Big)\Big(s^{kbc}+
\frac{1}{c^2}(u^b\Xi^{kc}-u^c\Xi^{kb})\Big) =
\\ \label{C3}
&=& \st{\td}{\Lambda}_c\frac{1}{c^2}\Xi^c +
\Lambda_c\st{\td}{u}_b\frac{1}{c^2}s^{bc}+
\Lambda_{c;k}\Xi^{kc}+\Lambda_cu_{b;k}s^{kbc}.\hspace{1cm}
\vspace{.3cm}\eey
Inserting \R{P13} and \R{C3} into \R{E7} results in
\bey\nonumber
S^k{_{;k}} &\equiv&
\frac{1}{c^2}\Big(\st{\td}{s}-\kappa\st{\td}{n}-
\lambda\st{\td}{e} -
\Lambda_c\st{\td}{\Xi}{^c}\Big) +
\\ \nonumber
&&+\ \frac{1}{c^2}\Big(s-\kappa n-\lambda e-
\Lambda_c \Xi^c-\lambda c^2p \Big)u^k{_{;k}} + 
\\ \nonumber
&& +\ \lambda_{,k} q^k +\frac{1}{c^2}\lambda\st{\td}{u}_mp^m +
\lambda u_{m;k}\pi^{km}+
\lambda u_m(G^m + k^m) + 
\\ \nonumber
&&+\ \Lambda_c\st{\td}{u}_b\frac{1}{c^2}s^{bc}+
\Lambda_{c;k}\Xi^{kc}+\Lambda_cu_{b;k}s^{kbc}+
\Lambda_cu_b(H^{bc}+m^{bc})+
\\ \label{aE7}
&&+\  \Big(s^k - \lambda q^k - \Lambda_c \Xi^{kc}\Big)_{;k}\
\equiv\ \sigma +\varphi.
\vspace{.3cm}\eey
As already mentioned, the entropy identity has to be transferred into the
expression for the entropy production by specifying the entropy flux $s^k$,
the entropy density $s$, the entropy supply $\varphi$ and the three for the
present arbitrary quantities $\kappa$, $\lambda$ and $\Lambda_c$.
\vspace{.3cm}\newline
Obviously, \R{aE7} contains terms of different kind: a divergence of a vector
perpendicular to $u^k$ (the last term of \R{aE7}), time derivatives of intensive
quantities (the first row of \R{aE7}), two terms stemming from the
field equations (last terms of the third and fourth row of \R{aE7}), 
three terms containing spin (3+1)-components
(in the fourth row of \R{aE7}) and three further terms containing
(3+1)-components of the energy-momentum tensor (in the third row
of \R{aE7}). This structure of the entropy identity allows to choose a state space
and by virtue of it, to define the entropy density, the
entropy supply, the entropy flux, the gr-Gibbs equation
and the gr-Gibbs-Duhem equation which all are represented
in the next sections.

\subsection{The entropy supply}

If the system under consideration is isolated, the external sources vanish
\bee{E10} 
k^m\ \doteq\ 0,\qquad m^{bc}\ \doteq\ 0,
\ee
and with them also the {\em entropy supply}
\bee{P22}
\varphi\ \equiv\ 0.
\ee
Thus, because the entropy supply is generated by external sources, 
we define
\bee{C8}
\varphi\ :=\ \lambda u_m k^m+\Lambda_cu_bm^{bc}.
\ee

\subsection{State space, gr-Gibbs equation and entropy flux \label{GE}}

We now choose a {\em state space} which belongs to a one-component
spin system in local equilibrium\footnote{Local equilibrium means:
The state at each event is described by a set of equilibrium variables which
change from event to event generating gradients of equilibrium variables causing irreversible processes.} and which is spanned by the particle number $n$,
the energy density $e$ and the spin density vector $\Xi_c$
\bee{L6k1}
\boxplus\ =\ (n, e, \Xi_c).
\ee
According to \R{E3} and \R{E4}, the 3-indexed spin is only partly taken
into account, namely by $\Xi_c$ and $\Xi^{kc}$. Here, $\Xi_c$ is an
independent state variable, whereas $\Xi^{kc}$ represents a constitutive
property according to \R{L7a}.
\vspace{.3cm}\newline
The {\em gr-Gibbs equation} is given by the relativistic time derivative of
the entropy density $s$ which is composed of time derivatives belonging
to the chosen state space. Such time derivatives appear only in the first
row of \R{aE7}\footnote{The acceleration $\st{\td}{u}_m$ is not
a material property, but one of the kinematical invariants.}. Consequently, we define
\bee{G2}
\st{\td}{s}\ :=\ \kappa\st{\td}{n}+\lambda\st{\td}{e} +
\Lambda_c\st{\td}{\Xi}{^c}.
\vspace{.3cm}\ee
Up to here, the quantities $\kappa,\ \lambda,\ \Lambda_c$ introduced into
the entropy identity \R{aE5} are unspecified. Taking the gr-Gibbs 
equation \R{G2} into consideration, such a specification is now possible:
$\lambda$ is the reciprocal rest-temperature
\bee{aL7a}
\lambda\ :=\ \frac{1}{T},
\ee
$\kappa$ is proportinal to the chemical potential
\bee{bL7a}
\kappa\ :=\ -\frac{\mu}{T},
\ee
and $\Lambda_c$ is analogous to \R{bL7a} proportional to a spin potential
\bee{cL7a}
\Lambda_c\\ :=\ -\frac{\mu_c}{T}.
\ee
These quantities as all the others which do not belong to the state space
variables \R{L6k1} are constitutive quantities describing the material by
constitutive equations. These constitutive quantities are
\bee{L7a}
{\bf M}\ =\ (T,\mu,\mu_c,p^k,q^k,p,\pi^{km},s^{km},s^{ckm},\Xi^{km}).
\ee
They all --including the entropy density $s$ and the entropy flux density $s^k$-- are
functions of the state space variables
\bee{L7b}
{\bf M}\ =\ {\cal M}(\boxplus).
\ee
These {\em constitutive equations}\footnote{How to use the constitutive
equations in connection with the gravitational field equations \R{b5a} see
\C{MUBOMAU}.} are out of scope of this paper.
\vspace{.3cm}\newline
Because of \R{aL7a}, the term $\lambda q^k$ is as in CT a part of the
entropy flux. Consequently, we define the {\em entropy flux} according to
the last row of \R{aE7}
\bee{G1}
s^k\ :=\ \lambda q^k +\Lambda_c \Xi^{kc}.
\ee
According to \R{L7a}, the entropy flux density is also a constitutive
quantity.

\subsection{Entropy density and gr-Gibbs-Duhem equation}

According to the second row of the entropy identity \R{aE7},
we define the {\em entropy density}
\bee{G2a}
s\ :=\ \kappa n+\lambda e+\Lambda_c \Xi^c+\lambda c^2p.
\ee
This definition has to be in accordance with the gr-Gibbs equation
\R{G2}. As usual in non-relativistic thermostatics, we demand
a {\em gr-Gibbs-Duhem equation} of the intensive variables
\bee{G2c}
\st{\td}{\kappa}n+\st{\td}{\lambda}(e+c^2p)+ \st{\td}{\Lambda}_c\Xi^c+
\lambda c^2\st{\td}{p}\ =\ 0. 
\ee

\subsection{The entropy production}

Inserting the entropy supply \R{C8}, the entropy flux \R{G1}, the
gr-Gibbs equation \R{G2} and the entropy density \R{G2a} into the
entropy identity \R{aE7}, we obtain the {\em entropy production}
\bey\nonumber
\sigma&=&
\ \lambda_{,k} q^k +\frac{1}{c^2}\lambda\st{\td}{u}_mp^m +
\lambda u_{m;k}\pi^{km}+
\lambda u_mG^m+ 
\\ \label{C9a}
&&+\ \Lambda_c\st{\td}{u}_b\frac{1}{c^2}s^{bc}+
\Lambda_{c;k}\Xi^{kc}+\Lambda_cu_{b;k}s^{kbc}+
\Lambda_cu_bH^{bc}.
\vspace{.3cm}\eey
We get by taking \R{aP2c}$_1$ and \R{P6}$_3$ into account
\bee{C9c}
\frac{1}{c^2}\lambda\st{\td}{u}_mp^m +
\Lambda_c\st{\td}{u}_b\frac{1}{c^2}s^{bc}\ =\ 
-\ \frac{1}{c^2}u_m\Big(\lambda\st{\td}{p}{^m}+
\Lambda_c\st{\td}{s}{^{mc}}\Big). 
\ee
Putting together
\bey\nonumber
-\ \frac{1}{c^2}u_m\Big(\lambda\st{\td}{p}{^m}+
\Lambda_c\st{\td}{s}{^{mc}}\Big) +\lambda u_mG^m +
\Lambda_cu_bH^{bc}\ =\ 
\\ \label{C9d}
=\ u_m\Big[\lambda\Big(G^m-\frac{1}{c^2}\st{\td}{p}{^m}\Big) +
\Lambda_c\Big(H^{mc}-\frac{1}{c^2}\st{\td}{s}{^{mc}}\Big)\Big],
\eey
the entropy production \R{C9a} results in
\bey\nonumber
\sigma &=&
\ \lambda_{,k} q^k + \Lambda_{c;k}\Xi^{kc} +
u_{m;k}\Big(\lambda \pi^{km} + \Lambda_cs^{kmc}\Big)+
\\ \label{C9e}
&&+\ u_m\Big[\lambda\Big(G^m-\frac{1}{c^2}\st{\td}{p}{^m}\Big) +
\Lambda_c\Big(H^{mc}-\frac{1}{c^2}\st{\td}{s}{^{mc}}\Big)\Big],
\eey
an expression which belongs to a general-covariant one-component spin system.
The entropy production depends on the Geo-SMEC-terms of the balance equations,
that means, the same material has different entropy productions in 
space-times of different theories. Entropy flux and -density, gr-Gibbs and
gr-Gibbs-Duhem equation do not depend on Geo-SMEC-terms because
energy-momentum and spin tensor are independent of the
Geo-SMEC-terms\footnote{That is obvious, because the (3+1)-splits \R{P2c}
and \R{L15b} are valid for all space-times.}.

\section{Further Forms of Entropy Production}

The gradient of the velocity can be decomposed into its kinematical
in\-va\-riants: symmetric traceless shear $\sigma_{nm}$, expansion $\Theta$, 
anti-symmetric rotation $\omega_{nm}$ and acceleration $\st{\td}{u}_n$ \C{ELL}
\byy{aP27c}
u_{l;k}\ =\ \sigma_{lk} + \omega_{lk} + \Theta h_{lk}
+\frac{1}{c^2}\st{\td}{u}_l u_k,\hspace{2.cm}
\\ \label{P28}
\sigma_{lk}=\sigma_{kl},\ \omega_{lk}=-\omega_{kl},\quad 
u^l\sigma_{lk}=\sigma_{lk}u^k=u^l\omega_{lk}=\omega_{lk}u^k=0,
\\ \label{P29}
\sigma^k_{k}=\omega^k_{k}=0,\quad \Theta:=u^k_{k}.\hspace{1.8cm}
\eey
Consequently, the third term of \R{C9a} can be replaced by
\bee{E13}
\lambda u_{m;k}\pi^{km}\ =\ \lambda\sigma_{mk}\pi^{(km)} +
\lambda\omega_{mk}\pi^{[km]}.
\vspace{.3cm}\ee
We now derive another shape of the entropy production: Starting out
with the entropy identity \R{E6}, we obtain the entropy production by
taking the entropy flux \R{G1}, the entropy density \R{G2a} and the
energy-momentum balance \R{b5a}$_1$, the particle balance \R{P3}$_1$
and \R{P8}$_1$
into account. Inserting these quantities, we obtain for isolated systems
\bee{E13a}
\sigma\ =\ \frac{1}{c^2}\Big[\lambda c^2 pu^k\Big]_{;k} +
\st{\td}{\kappa}\frac{1}{c^2}n +
(\lambda u_m)_{;k}T^{km} + \lambda u_mG^m +
\Big(\Lambda_{c}u_b S^{kbc} \Big)_{;k},
\ee
and the first term of \R{E13a} is 
\bee{aE13a}
\frac{1}{c^2}\Big[\lambda c^2 pu^k\Big]_{;k}\ =\ 
(\lambda p)^\td + \lambda p u^k{_{;k}}.
\vspace{.3cm}\ee
For the sequel, we need an additional expression whose validity is independent
of the entropy production because it represents an identity
\bey\nonumber
(\lambda u_{m})_{;k}\Big(\frac{1}{c^4}eu^ku^m -ph^{km}\Big)&=&
(\lambda_{,k}u_m + \lambda u_{m;k})
\Big(\frac{1}{c^4}eu^ku^m -ph^{km}\Big)\ =\
\\ \label{dG2}
&=&
\frac{e}{c^2}\lambda^\td - \lambda pu^k{_{;k}}.
\vspace{.3cm}\eey
The sum of \R{E13a} and \R{dG2} results in
\bey\nonumber
\sigma &=&
(\lambda p)^\td + \frac{e}{c^2}\lambda^\td +\st{\td}{\kappa}\frac{1}{c^2}n +
\\ \nonumber
&&+(\lambda u_m)_{;k}\Big(T^{km}-\frac{1}{c^4}eu^ku^m +ph^{km}\Big)+
\\ \label{eG2}
&&+\lambda u_mG^m +\Big(\Lambda_{c}u_b\Big)_{;k}S^{kbc}+
\Lambda_{c}u_bH^{bc}.
\eey
Replacing the first two terms by the gr-Gibbs-Duhem equation \R{G2c},
we obtain by taking \R{C3} into account
\bey\nonumber
\sigma &=&
-\frac{1}{c^2}\st{\td}{\Lambda}_c\Xi^c +
(\lambda u_m)_{;k}\Big(T^{km}-\frac{1}{c^4}eu^ku^m +ph^{km}\Big)+
\\ \label{eG2a1}
&&+\ \lambda u_mG^m +\Big(\Lambda_{c}u_b\Big)_{;k}S^{kbc}+
\Lambda_{c}u_bH^{bc}\ =
\\ \nonumber
&=& (\lambda u_m)_{;k}\Big(T^{km}-\frac{1}{c^4}eu^ku^m +ph^{km}\Big)+
 \lambda u_mG^m +
\\ \label{eG2a}
&&+\ \Lambda_{c}u_b\Big(H^{bc}-
\frac{1}{c^2}\st{\td}{s}{^{bc}}\Big) +
\Lambda_{c;k}\Xi^{kc}+\Lambda_cu_{b;k}s^{kbc}.
\vspace{.3cm}\eey
Here in contrast to \R{C9e}, the heat flux $q^k$ and the friction
tensor $\pi^{km}$ do not appear. They are replaced by the first term of 
\R{eG2a} describing the deviation of the material from a perfect one.
The second row represents the influence of the chosen state space \R{L6k1}
on the entropy production introduced by the third term of \R{aE5}.
\vspace{.3cm}\newline
We now consider the thermodynamical results: if we start out with the entropy
identity for which third term of \R{aE5} is set to zero $\Lambda_c\equiv 0$ \C{SL}.
That means, we change the state space \R{L6k1} into
\bee{dG2a}
\boxplus_0\ =\ (n, e).
\ee
The entropy production \R{eG2a} becomes
\bee{eG2b}
\sigma_0\ =\
(\lambda u_m)_{;k}\Big(T^{km}-\frac{1}{c^4}eu^ku^m +ph^{km}\Big)+
\lambda u_mG^m,
\ee
and the entropy flux \R{G1} is
\bee{C16a}
s^k_0\ =\ \lambda q^k .
\ee
The gr-Gibbs equation \R{G2} becomes
\bee{C17a}
\st{\td}{s}_0\ =\ \kappa\st{\td}{n} + \lambda\st{\td}{e},
\ee
and the entropy density \R{G2a} is
\bee{C18a}
s_0\ =\ \kappa n +\lambda e + c^2\lambda p.
\ee
Finally, the gr-Gibbs-Duhem equation \R{G2c} results in
\bee{C19a}
\st{\td}{\kappa}n + \st{\td}{\lambda}\Big(e+c^2p\Big)+
\st{\td}{p}c^2\lambda\ =\ 0.
\vspace{.3cm}\ee
The thermodynamical quantities \R{eG2b} to \R{C19a} base on the chosen
state space as a comparison with \R{G2}, \R{L6k1}, \R{G2a} and \R{G2c}
demonstrates. Consequently, the thermodynamical quantities are not "absolute",
they belong to a thermodynamical scheme implemented by a chosen state space.
The spin does not appear in the thermodynamical quantities \R{eG2b} to \R{C19a}
in contrast to \R{G2}, \R{L6k1}, \R{G2a} and \R{G2c}. Also the regard of the spin
balance \R{b5a}$_3$ is different: the entropy production \R{eG2a} takes it
explicitly into account, whereas spin parts do not appear in \R{eG2b}
\C{ST}. Here, the spin is a constitutive quantity
and not a state space variable. For the sequel, we use the state space
\R{L6k1} because of its generality and consider the restricted state space \R{dG2a}
as a special case.
\vspace{.3cm}\newline
The special case \R{dG2a} 
can be easily obtained by the setting $\Lambda_c\doteq 0$. Especially
in GR, the energy-momentum tensor is symmetric and the external sources and
the Geo-SMEC-terms are zero. Thus \R{eG2a} results in
\bee{C19b}
\sigma_0^{GR}\ =\
\frac{1}{2}\Big[(\lambda u_m)_{;k} + (\lambda u_k)_{;m}\Big]
\Big(T^{km}-\frac{1}{c^4}eu^ku^m +ph^{km}\Big).
\ee
Consequently, the entropy production vanishes in GR for perfect materials
or/and if the spacetime allows that the 4-temperature vector is a Killing field. 
This is the well-known result derived  in \C{ST} which is here worked out using
a more general aspect. More details in connection with equilibrium will be
discussed in sect.\ref{EC}.

\section{Non-dissipative Materials\label{NDM}}

As already mentioned, the entropy production is always connected with
a chosen state space. Thus \R{eG2a} belongs to \R{L6k1}, and \R{eG2b}
to \R{dG2a}: the entropy production depends
on the material and on the space-time, a statement which is also valid for
its zero. A {\em non-dissipative material} is characterized by vanishing entropy
production \R{eG2a} even in the case of non-equilibrium\footnote{Vanishing
entropy production is necessary, but not sufficient for equilibrium.} 
independently of the specially chosen space-time. 
Consequently by de\-fi\-nition, all processes of non-dissipative materials are
reversible\footnote{Reversible "processes" are trajectories in the state space
consisting of equilibrium states.}, and therefore these materials are
those of thermostatics. If the state space is changed, it may be that a
non-dissipative material becomes dissipative.
\vspace{.3cm}\newline
Starting out with \R{eG2a}, we point out a set of conditions which is sufficient
that a material is non-dissipative, that means, its entropy production vanishes
independently of the space-time. These conditions are generated by 
setting individual terms in \R{eG2a} to zero. First the non-dissipative material
is perfect
\bee{U0b}
T^{kl}_{\sf ndiss}\ \doteq\ \frac{1}{c^4}eu^ku^l - ph^{kl}\ ,\quad
\longrightarrow\quad
T^{[kl]}_{\sf ndiss}\ =\ 0, 
\ee
for which the first term of \R{eG2a} vanishes. Also the second term must
vanish
\bee{aU0}
G^m_{\sf ndiss}\ \doteq\ 0,
\ee
that means, the Geo-SMEC-term of the energy-momentum balance
has to be zero. 
\vspace{.3cm}\newline
The second row of \R{eG2a} depends on $\Lambda_c$
which introduces according to the last term of \R{aE5} a spin part explicitly
into the state space \R{L6k1}. There are now two possibilities for vanishing the
second row of \R{eG2a}: first
\bee{U1}
H^{bc}_{\sf ndiss}\ =\ 0,\qquad \st{\td}{s}_{\sf ndiss}^{bc}\ =\ 0,\qquad
\Psi^{kbc}_{\sf ndiss}\ =\ 0,
\ee
or second
\bee{U2}
\Lambda_c^{\sf ndiss}\ =\ 0.
\ee
If now $\Lambda_c$ vanishes in 
\R{eG2a} --and consequently also in \R{aE5}-- spin terms cannot appear
in the state space, with the result that we have to choose the state space
\R{dG2a} instead of \R{L6k1}.
\vspace{.3cm}\newline
Finally, we proved two statements which presupposes different state spaces 
for non-dissipative materials.
\vspace{.3cm}\newline
$\blacksquare$
{\sf Proposition I:} The five altogether sufficient conditions characterizing
non-dissipative materials are: (i) the material is perfect, (ii) the Geo-SMEC-terms
of the energy-momentum and of the spin balance vanish, (iii) the spin is
$S^{kbc}_{\sf ndiss}=u^k\Phi^{bc}_{\sf ndiss}$,
(iv) the spin density \R{U1}$_2$ is covariantly constant and
(v) the state space is spanned by the particle number, the energy density and
the spin density vector $\Xi_c$.
$\mbox{}$\hfill $\blacksquare$
\vspace{.3cm}\newline
The second proposition presupposing the state space \R{dG2a} is as follows:
\vspace{.3cm}\newline
$\blacksquare$
{\sf Proposition II:}
The three altogether sufficient conditions characterizing
non-dissipative materials are: (i) the material is perfect, (ii) the Geo-SMEC-term
of the energy-momentum balance vanishes, (iii) the state space is spanned by the
particle number and the energy density.\hfill $\blacksquare$

\section{Equilibrium\label{EQ}}
\subsection{Equilibrium Conditions\label{EC}}

We start out with the question:
How are equilibrium and non-dissipative materials
related to each other ? Concerning non-dissipative materials, we are looking
for material properties enforcing vanishing entropy production for all admissible
space-times. Concerning equilibria, we are asking for space-times in which
materials can be at equilibrium. This is defined by {\em equilibrium
conditions} which are divided into necessary and supplementary ones \C{B}.
The necessary ones are given by vanishing entropy production and vanishing
entropy flux density
\bee{P23} 
\sigma^{eq}\ \doteq\ 0\quad\wedge\quad s^k_{eq}\ \doteq\ 0.
\ee
Supplementary equilibrium conditions are given by vanishing material time 
derivatives, except that of the 4-velocity
\bee{P24}
\boxplus^\bullet_{eq}\ \doteq\ 0,\qquad\boxplus\
\neq\ u^l.
\ee
Consequently according to
\R{G2} and \R{G2c}, the gr-Gibbs and the gr-Gibbs-Duhem equations
are identically satisfied in equilibrium.
Whereas in sect.\ref{NDM} the non-dissipative materials are defined
independently of the admissible space-times, here a material
in a given space-time is considered and the equilibrium conditions \R{P23} and \R{P24} are valid. 
\vspace{.3cm}\newline
From \R{P3}$_1$ follows
\bee{L1c}
\st{\td}{n}\ :=\ n_{,k}u^k\ =\ -nu^k{_{;k}}\quad\longrightarrow\quad
u^k{_{;k}}\ =\ -\frac{\st{\td}{n}}{n}.
\ee
According to \R{L1c}$_3$ and \R{P24}$_1$, the divergence of the 4-velocity
--that is the expansion \R{P29}$_2$-- vanishes always in
equilibrium\footnote{for arbitrary space-times and materials}
\bee{P25c}
u^k{_{;k}}{^{eq}}\ =\ 0.
\vspace{.3cm}\ee
Starting out with the identity \R{dG2} and taking \R{P24} and \R{P25c}
into account, we obtain for all equilibria
\bee{aP25c}
(\lambda u_{m})^{eq}_{;k}\Big(\frac{1}{c^4}eu^ku^m -
ph^{km}\Big)^{eq}\ =\ 0.
\ee
Because the second bracket of \R{aP25c} is never zero, 
the 4-temperature vector $\lambda u_m$ is independently of the material
a Killing vector in equilibrium
\bee{aP25d}
\Big[(\lambda u_m)_{;k} + (\lambda u_k)_{;m}\Big]^{eq}\ =\ 0.
\ee
The equilibrium conditions \R{P25c} and \R{aP25d} are induced by 
\R{P24}$_1$ independently of the entropy production and the material.
No equilibria are possible
in space-times which do not allow the validity of \R{P25c} or/and \R{aP25d}. 
Obvious is that the conditions \R{P25c} and \R{aP25d} are necessary,
but not sufficient for equilibrium, because they do not guarantee vanishing
entropy production \R{eG2a} or \R{eG2b}, except for the case of GR
according to \R{C19b}. Hence, vanishing entropy production in GR means
two different things: the system may be in equilibrium\footnote{if all the other
equilibrium conditions are valid} or the system is non-dissipative and reversible
processes occur.
\vspace{.3cm}\newline
The expression of the entropy production \R{C9e}
becomes in equilibrium by taking \R{E13} into account 
\bey\nonumber
0 &=& \lambda_{,k}^{eq}q^k_{eq} +
\Lambda_{c;k}^{eq}\Xi^{kc}_{eq}+
\\ \nonumber
&&+\ \sigma_{mk}^{eq}\Big(\lambda^{eq}\pi^{(km)}_{eq} +
\Lambda_c^{eq}s^{(km)c}_{eq}\Big)+
\omega_{mk}^{eq}\Big(\lambda^{eq}\pi^{[km]}_{eq} +
\Lambda_c^{eq}s^{[km]c}_{eq}\Big)+
\\ \label{P28a}
&&+\ u_m^{eq}\Big(\lambda^{eq} G^m_{eq} +
\Lambda_c^{eq}H^{mc}_{eq}\Big).
\eey
From \R{G1} and \R{P23}$_2$ follows
\bee{P31}
q^k_{eq}\ =\ -\frac{1}{\lambda^{eq}}
\Lambda_c^{eq}\Xi^{kc}_{eq}\quad\longrightarrow\quad
\lambda_{,k}^{eq}q^k_{eq}\ =\ 
-\frac{\lambda_{,k}^{eq}}{\lambda^{eq}}
\Lambda_c^{eq}\Xi^{kc}_{eq}.
\ee
Inserting \R{P31} into \R{P28} results in
\bey\nonumber
0 &=&\Xi^{kc}_{eq}\Big(\Lambda_{c;k}^{eq}-
\frac{\lambda_{,k}^{eq}}{\lambda^{eq}}\Lambda_c^{eq}\Big)+
\\ \nonumber
&&+\ \sigma_{mk}^{eq}\Big(\lambda^{eq}\pi^{(km)}_{eq} +
\Lambda_c^{eq}s^{(km)c}_{eq}\Big)+
\omega_{mk}^{eq}\Big(\lambda^{eq}\pi^{[km]}_{eq} +
\Lambda_c^{eq}s^{[km]c}_{eq}\Big)+
\\ \label{P32}
&&+\ u_m^{eq}\Big(\lambda^{eq} G^m_{eq} +
\Lambda_c^{eq}H^{mc}_{eq}\Big).
\vspace{.3cm}\eey
In contrast to the material-independent equilibrium conditions \R{P25c}
and \R{aP25d}, the equilibrium conditions \R{P28a} to \R{P32} depends on material
and space-time. Each of these three condition is necessary for equilibrium,
and altogether they are sufficient for equilibrium because the field equations
\R{b5a} and the entropy supply \R{C8} are taken into account.
\vspace{.3cm}\newline
An other shape of \R{P32} can be derived from \R{eG2a} by taking \R{aP25d}
into account
\bey\nonumber
0 &=&(\lambda u_m)_{;k}^{eq}T^{[km]}_{eq}+
\\ \label{P29a}
&&+\ \Lambda_{c;k}^{eq}\Xi^{kc}_{eq}+
\Lambda_c^{eq}u_{m;k}^{eq}s^{kmc}_{eq}+
 u_m^{eq}\Big(\lambda^{eq}G^m_{eq} +
\Lambda_c^{eq}H^{mc}_{eq}\Big).
\vspace{.3cm}\eey
This equilibrium condition is satisfied in GR because the energy-momentum tensor
is symmetric, both Geo-SMEC-term vanishes and the state space \R{dG2a} is used.
General solutions of \R{P32} and \R{P29a}, that means, to find all
couples --material $\leftrightarrow$ space time-- which satisfy
\R{P32} and \R{P29a}, cannot be achieved. Therefore, we discuss some
special cases of equilibria in the next section.

\subsection{Special equilibria}

We now decompose the equilibrium condition \R{P32} into a set of terms
representing special cases of equilibria and
which altogether enforce the validity of \R{P32}:
\byy{P33}
\Xi^{kc}_{eq}\Big(\Lambda_{c;k}^{eq}-
\frac{\lambda_{,k}^{eq}}{\lambda^{eq}}\Lambda_c^{eq}\Big)\ \doteq\ 0,
\\ \label{P34}
\sigma_{mk}^{eq}\Big(\lambda^{eq}\pi^{(km)}_{eq} +
\Lambda_c^{eq}s^{(km)c}_{eq}\Big)\ \doteq\ 0,
\\ \label{P35}
\omega_{mk}^{eq}\Big(\lambda^{eq}\pi^{[km]}_{eq} +
\Lambda_c^{eq}s^{[km]c}_{eq}\Big)\ \doteq\ 0,
\\ \label{P36}
 u_m^{eq}\Big(\lambda^{eq} G^m_{eq} +
\Lambda_c^{eq}H^{mc}_{eq}\Big)\ \doteq\ 0.
\eey
If we do not restrict the spin material under consideration, the bracket
in \R{P33} has to be zero resulting in a differential
equation for the spin potential \R{cL7a}
\bee{P37}
\lambda^{eq}\Lambda_{c;k}^{eq}-
\lambda_{,k}^{eq}\Lambda_c^{eq}\ =\ 0.
\ee
Because this equilibrium condition is pretty exotic, we restrict our
discussion to spin materials with vanishing $\Xi^{kc}$. According to
\R{L15b} and \R{P31}, we obtain
\bee{P38}
\Xi^{kc}\ \doteq\ 0,\quad\longrightarrow\quad
S^{kbc}\ =\ u^k\Phi^{bc} + s^{kbc}\quad\wedge\quad
q_k^{eq}\ =\ 0.
\ee
According to \R{P34} and \R{P35}, the couple stress $s^{kbc}$
modifies the friction tensor
\bee{P39}
\Pi^{km}_{eq}\ :=\ 
\lambda^{eq}\pi^{km}_{eq} +
\Lambda_c^{eq}s^{kmc}_{eq},
\ee
and necessary material-independent equilibrium conditions are
\bee{P39a}
\sigma_{mk}^{eq}\ =\ 0,\qquad\omega_{mk}^{eq}\ =\ 0.
\ee
Another necessary equilibrium condition is \R{P36}. In connexion with
\R{P29a}, we obtain
\bee{P42}
(\lambda u_m)_{;k}^{eq}T^{[km]}_{eq}+
\Lambda_c^{eq}u_{m;k}^{eq}s^{kmc}_{eq}\ =\ 0,
\ee
a relation which is satisfied, if we have e.g.
\bee{P42a}
(\lambda u_m)_{;k}^{eq}\ =\ 0\quad\wedge\quad s^{kmc}_{eq}\ =\ 0.
\vspace{.3cm}\ee
Using the state space \R{dG2a}, \R{P33} to \R{P36} result in \R{P39a} or
\bee{P42b}
\pi_{eq}\ =\ 0\quad\wedge\quad u_m^{eq}G^m_{eq}\ =\ 0.
\ee

\subsection{Frenkel materials}

Materials defined by the special spin\footnote{taking \R{L15c}$_3$into account}
\bee{P43}
\Psi^{kbc}_{\sf FR}\ \doteq\ 0\quad\wedge\quad
\Big[\Phi^{bc}_{\sf FR}u_b\ \doteq\ 0
\quad\longrightarrow\quad\Xi^c_{\sf FR}\ =\ 0\Big]
\ee
are called {\em Frenkel materials}.
According to \R{P43}$_2$, Frenkel materials belong to the state space \R{dG2a},
and their spin is
\bee{P43a} 
S^{kbc}_{\sf FR}\ =\ u^ks^{bc}.
\ee
According to \R{eG2a}, a necessary equilibrium condition of Frenkel materials is
\bee{P43b}
0\ =\ (\lambda u_m)_{;k}^{eq}\Big(T^{km}_{\sf FR}-
\frac{1}{c^4}e_{\sf FR}u^ku^m +ph^{km}_{\sf FR}\Big)^{eq}+
u_m^{eq}\lambda^{eq} G^m_{{\sf FR}eq}.
\ee
If the Frenkel material is dissipative, the equilibrium conditions are
\R{P25c}, \R{aP25d} and according to \R{P43b}
\bee{P43c}
(\lambda u_m)_{;k}^{eq}T^{[km]}_{{\sf FR}eq}\ =\ 0,\qquad
u_m^{eq}\lambda^{eq} G^m_{{\sf FR}eq}\ =\ 0,
\ee
and additionally according to \R{P31}$_1$ and \R{P34}/\R{P35}
\bee{P43d}
q^k_{{\sf FR}eq}\ =\ 0,
\qquad\sigma_{mk}^{eq}\lambda^{eq}\pi^{(km)}_{{\sf FR}eq}\ =\ 0,
\qquad\omega_{mk}^{eq}\lambda^{eq}\pi^{[km]}_{{\sf FR}eq}\ =\ 0.
\ee

\section{Discussion}

Starting out with the entropy identity derived in \C{B} and specifying
entropy flux, entropy density and entropy supply, different expressions of the entropy
production in general-relativistic space-times are determined by taking
the gr-Gibbs and the gr-Gibbs-Duhem equations into regard. All these
thermodynamical quantities depend on the chosen state space which
in general is more extended than that of General Relativity. Beyond that,
the entropy production of a general-covariant one-component spin
system depends on the so-called Geo-SMEC-terms which are located at
the rhs of the balance equations, thus discriminating between different
general-covariant theories. 
\vspace{.3cm}\newline
Well-known relations of General Relativity are generalized for theories
based on post-Riemannian space-times.
In this case, the interrelation between geometric and constitutive
quantities in the expression for the entropy production becomes more
complex. Consequently, the zero of the entropy production can be realized
by a variety of conditions imposed on constitutive and/or geometric quantities. 
One condition of them is the fact that the entropy production vanishes
for perfect materials, if the state space does not include spin terms and
if the Geo-SMEC-term of the energy-momentum balance is zero. That is
just the well-known case of Ge\-ne\-ral Relativity. 
\vspace{.3cm}\newline
Vanishing of entropy production is only necessary, but not sufficient
for equi\-li\-brium. This necessary condition has to be complemented by
"supplementary equilibrium conditions" for describing equilibrium
sufficiently. Two supplementary equilibrium conditions are independent
of the entropy production restricting the space-time independently of
the material: the expansion va\-nishes and the 4-temperature
vector is a Killing field. Equilibria are impossible, if one of these conditions
is not satisfied.
\vspace{.3cm}\newline
From the viewpoint of material theory, the conditions are interesting
for which the entropy production vanishes whatever the properties of
the space-time of the considered theory may be. One set of conditions
for non-dissipativity is:
the material is perfect, the Geo-SMEC-term of the energy-momentum
balance vanishes and the state space is spanned by particle number
and energy density. That is again the special case of General Relativity.

\end{document}